\documentclass[preprint]{aastex}
\usepackage{emulateapj5}
\usepackage{apjfonts}
\usepackage{natbib}
\usepackage{epsfig}
\input psfig.tex

\def\aa{{A\&A}}

\def\aj{{AJ}}

\def\annrev{{ARA\&A}}
\def\apj{{ApJ}}
\def\apjs{{ApJS}}

\def\etal{{et al. }}

\def\hal{H$\alpha$}
\def\hb{H$\beta$}

\def\kms{km s$^{-1}$}
\def\lamb{$\lambda$}
\def\lax{{$\mathrel{\hbox{\rlap{\hbox{\lower4pt\hbox{$\sim$}}}\hbox{$<$}}}$}}
\def\gax{{$\mathrel{\hbox{\rlap{\hbox{\lower4pt\hbox{$\sim$}}}\hbox{$>$}}}$}}
\def\simlt{\lower.5ex\hbox{$\; \buildrel < \over \sim \;$}}
\def\simgt{\lower.5ex\hbox{$\; \buildrel > \over \sim \;$}}
\def\lum{ergs s$^{-1}$}

\def\mnras{{MNRAS}}

\def\pasp{{PASP}}

\def\percm2{cm$^{-2}$}
\def\peryr{yr$^{-1}$}

\def\solmass{$M_\odot$}
\def\oii{[\ion{O}{2}]}

\def\hii{\ion{H}{2}}
\def\oiii{[\ion{O}{3}]}

\def\oi{[\ion{O}{1}]}
\def\nii{[\ion{N}{2}]}

\def\sii{[\ion{S}{2}]}

\slugcomment{To Appear in {\it The Astrophysical Journal}.}
\shorttitle{STAR FORMATION IN AGNs}
\shortauthors{KIM, HO \& IM}

\begin{document}

\title{Constraints on the Star Formation Rate in Active Galaxies}
\author{Minjin Kim}
\affil{Astronomy Program, SEES, Seoul National University,
Seoul 151-742, Korea}
\author{Luis C. Ho}
\affil{The Observatories of the Carnegie Institution of
Washington, 813 Santa Barbara St., Pasadena, CA 91101}
\author{Myungshin Im}
\affil{Astronomy Program, SEES, Seoul National University,
Seoul 151-742, Korea}
\begin{abstract}
The \oii\ \lamb3727 emission line is often used as an indicator of star 
formation rate in extragalactic surveys, and it can be an equally effective 
tracer of star formation in systems containing luminous active galactic 
nuclei (AGNs). In order to investigate the ongoing star formation rate of the 
host galaxies of AGNs, we measured the strength of \oii\ and other optical 
emission lines from a large sample ($\sim 3600$) of broad-line (Type 1) AGNs 
selected from the Sloan Digital Sky Survey.  We performed a set of 
photoionization calculations to help evaluate the relative contribution of 
stellar and nonstellar photoionization to the observed strength of \oii.  
Consistent with the recent study of Ho (2005), we find that the observed \oii\ 
emission can be explained entirely by photoionization from the AGN itself, 
with little or no additional contribution from \hii\ regions.  This indicates 
that the host galaxies of Type 1 AGNs experience very modest star formation 
concurrent with the optically active phase of the nucleus.  By contrast, we 
show that the sample of ``Type 2'' quasars selected from the Sloan Digital Sky 
Survey does exhibit substantially stronger \oii\ emission consistent 
with an elevated level of star formation, a result that presents a challenge 
to the simplest form of the AGN unification model.
\end{abstract}


\section{Introduction}

It is now almost universally accepted that massive black holes are common and 
that they play an important role in many facets of galaxy evolution (see 
reviews in Ho 2004).  Numerous theoretical models have been proposed to 
explain the strong observed correlations between black hole mass and bulge 
properties (Magorrian et al. 1998; Gebhardt et al. 2000; Ferrarese \& Merritt 
2000).  While everyone agrees that the growth of black holes must be closely 
linked with galaxy formation (e.g., Silk \& Rees 1998; Kauffmann \& Haehnelt 
2000; Begelman \& Nath 2005; Robertson et al.  2006), there is no consensus as 
to exactly how the two very different processes involved---accretion and 
star formation---are really coupled.  Are they well synchronized, or does 
one process precede the other, and if so, what is the time lag?  When the 
black hole is actively growing, does the feedback from the active galactic 
nucleus (AGN) actually trigger or inhibit star formation in the host galaxy?
These important issues are unlikely to be settled through theoretical 
speculations or numerical simulations alone.  Some empirical guidance from 
observations would be welcomed.

In recent years, there has been mounting evidence that AGN activity and 
starburst activity are often intermixed.  The most commonly cited example 
occurs in ultraluminous infrared galaxies, whose dominant energy source has 
been much debated (e.g., Genzel et al.  1998).  Detailed studies of individual 
active galaxies have revealed a significant population of young stars in a 
number of instances, either through stellar absorption features (e.g., Boisson 
et al. 2000; Canalizo \& Stockton 2000; Cid Fernandes et al. 2001) or colors 
(e.g., Jahnke et al. 2004).  Additional hints for a close connection between 
star formation and black hole accretion comes from the high degree of chemical 
enrichment in quasars as deduced from analysis of their spectra (Hamann et al. 
2004 and references therein). Lastly, the host galaxies of quasars, at least 
in the nearby Universe, possess a rich supply of both molecular gas (Scoville 
et al. 2003) and dust (Haas et al. 2003), and thus have the potential for 
sustaining ongoing or future star formation.

While there is little doubt that the above observations indicate that AGN 
activity and star formation do often coincide within the same galaxy, one of 
the main challenges in interpreting this evidence in a physical context lies 
in the difficulty of establishing a direct, {\it causal}\ connection between 
the two phenomena.  An important step forward was recently achieved.  From an 
analysis of a large sample of narrow-line (Type 2) AGNs selected from the 
Sloan Digital Sky Survey (SDSS; York et al. 2000), Kauffmann et al. (2003) 
showed that, unlike low-luminosity sources (Ho et al. 2003), more powerful 
AGNs often exhibit stellar absorption features indicative of young to 
intermediate-age stars.  Although more difficult to ascertain due to the 
strong contamination from the bright nucleus, Kauffmann et al. show that the 
spectral signatures for young stars seem to persist even among broad-line 
(Type 1) AGNs.\footnote{We will use the terms ``Type 1'' and ``Type 2'' AGNs 
to refer to broad-line and narrow-line objects, respectively.  Whenever 
possible, we will refrain from using the terms ``Seyferts'' or ``quasars,'' 
which, for historical reasons, carry artificial luminosity criteria.} Most 
crucially, they find, for their primary sample of Type 2 objects, that the 
fraction of young stars in the central region of the host galaxies increases 
with increasing AGN luminosity, providing, for the first time, tantalizing 
evidence for a causal link between star formation and accretion.  

It is important to recognize, however, that the SDSS results constrain only 
the {\it post-starburst}\ population, with ages $\sim 10^8-10^9$ yr.  There is 
no information on the presence of younger, ionizing stars (ages \lax$10^7$ yr). 
Since the Type 2 AGN sample was identified using standard optical emission-line 
ratios designed to isolate accretion-dominated sources, {\it by selection}\ 
it excludes any sources with a sizable contribution from \hii\ regions.
On the other hand, it is clearly of significance to constrain the ongoing star 
formation rate (SFR) in AGNs, in order to assess the extent to which the 
two processes are coeval.  AGN lifetimes are uncertain, but current best 
estimates range from $10^6$ to $10^8$ yr (Martini 2004), precisely in the 
unexplored regime of interest.

Estimating SFRs in AGNs presents significant complications, since the AGN, by 
definition, dominates the integrated output of the source.  Nearly all of the 
standard measures of SFRs employed for inactive galaxies (e.g., Kennicutt 
1998) are rendered useless by emission from the AGN itself.  There is, 
however, one important exception: the \oii\ \lamb3727 line.  As discussed by 
Ho (2005), \oii, an emission line commonly used in spectroscopic surveys to 
estimate SFRs in galaxies at redshifts $z$ \gax\ 0.4, should be an equally 
effective tracer of \hii\ regions in Seyferts and quasars.  In such
high-ionization active galaxies, the \oii\ emission intrinsic to the 
narrow-line region (NLR) of the AGN is both observed and predicted to be 
low (e.g., Ferland \& Osterbrock 1986; Ho et al. 1993b).  If high-ionization 
AGNs experience substantial levels of ongoing star formation, the integrated
contribution from \hii\ regions will boost the strength of the \oii\ line 
(compared to, say, \oiii\ \lamb 5007, which can be largely ascribed to the AGN 
itself).  From a survey of the existing literature on \oii\ measurements on 
quasars, Ho (2005) concludes that their SFRs are surprisingly modest.  
More intriguingly, for a subset of low-redshift quasars with available 
molecular gas measurements, Ho finds that their star formation efficiencies 
(SFR per unit gas mass) appear to be abnormally low compared to either 
normal or infrared-bright galaxies.  He suggests that this unusual behavior
may be a manifestation of AGN feedback suppressing star formation.

The results of Ho (2005) regarding SFRs in quasars were based largely on \oii\ 
strengths deduced from composite spectra (ensemble averages) constructed from 
quasar surveys.  While the conclusions are statistically robust, it would be 
desirable to revisit this problem from actual measurements of a large, 
well-defined sample of individual objects.  This is accomplished in this 
paper, where we make use of an extensive sample of Type 1 AGNs selected from 
the SDSS to investigate the properties not only of \oii, but also a number of 
other diagnostically important narrow optical emission lines.  We perform a 
new set of photoionization calculations to demonstrate that the observed 
strength of \oii\ emission is entirely consistent with standard AGN 
photoionization.  These results support and strengthen the conclusion of Ho 
(2005) that Type 1 AGNs experience very limited ongoing star formation.  By 
contrast, we show that Type 2 quasars exhibit markedly enhanced \oii\ 
emission, and hence presumably elevated SFRs.  This result seems to violate 
the basic orientation-dependent unification model for AGNs.

We adopt the following cosmological parameters: $H_0 = 100\,h = 71$ 
\kms~Mpc$^{-1}$, $\Omega_{\rm m} = 0.27$, and $\Omega_{\Lambda} = 0.75$ 
(Spergel et al. 2003).

\section{Sample Selection}
Our goal is to investigate the narrow-line spectrum of a large, homogeneous 
sample of broad-line (Type 1) AGNs.  We utilize the database from SDSS 
(Stoughton et al. 2002; Strauss 
\epsfig{file=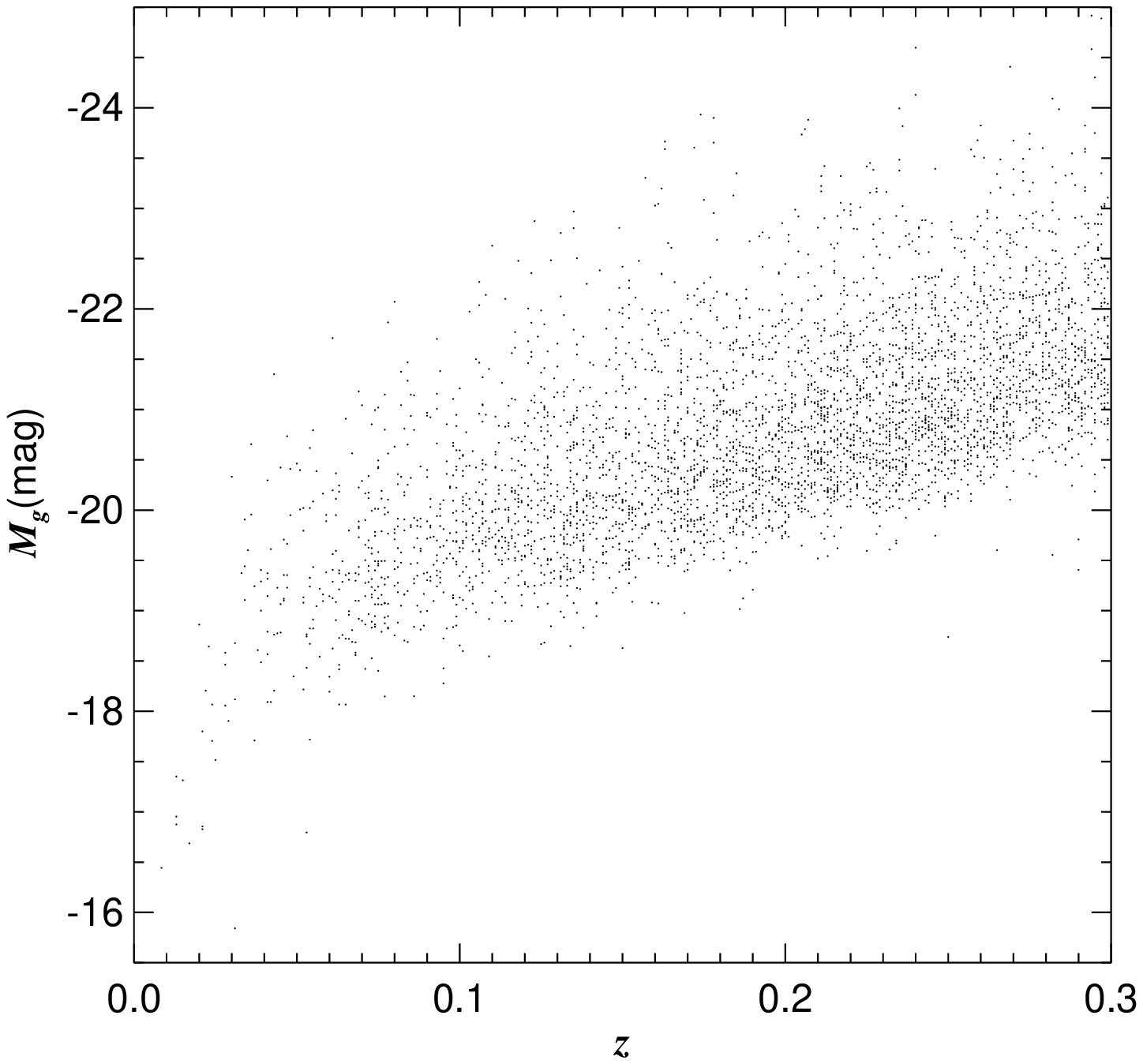,width=8.5cm,angle=0}
\vskip -5mm
\figcaption[fig1.ps]{
The distribution of \textit{g}-band absolute magnitudes
as a function of redshift for the sample. The magnitudes are ``PSF''
magnitudes, as defined by SDSS, and they have been corrected
for Galactic extinction.
}
\vskip 5mm
et al. 2002), whose  spectra
(taken with $3\arcsec$-diameter fibers) cover $\sim 3800 - 9200$ \AA\ with an
instrumental resolution of $\lambda/\Delta \lambda \approx 1800-2100$.  
This paper makes use of data from the Third Data Release (DR3) of SDSS 
(Abazajian \etal\ 2005), from which we have selected a total of 4086 
objects flagged by the spectroscopic pipeline as ``AGN'' that have a 
redshift confidence larger than 0.9 and a redshift $z < 0.3$.  The redshift cut
ensures that the diagnostically important \sii\ \lamb\lamb6716, 6731 doublet 
falls comfortably within the bandpass.  After inspection of the spectra, we 
discarded 498 that either had gaps in the spectral coverage or had at least 
one of the main emission lines corrupted, leaving a final sample of 3588 
objects.  Figure 1 shows the distribution of absolute (``PSF'') \textit{g}-band 
magnitudes as a function of redshift.  The magnitudes were corrected for 
Galactic extinction, using the extinction values of Schlegel et al. (1998) 
and the extinction curve of Cardelli et al. (1989), but we did not apply a 
K-correction because of the ambiguity introduced by the uncertain amount of 
galaxy contamination.  Since local field galaxies have an absolute $g$-band
magnitude of  $M_{*} = -20.44-5\log h \approx -21.2$ (Blanton et al. 2003), 
it is clear that some of our objects are significantly contaminated by host 
galaxy emission.   We simulated the expected magnitude due to K-correction by 
adding different relative fractions of quasar and galaxy light, which we 
approximate, respectively, using a composite quasar spectrum (Vanden Berk 
et al. 2001) and an early-type galaxy template (Coleman et al. 1980).   For an 
AGN contribution of 40\% or more at 5100 \AA, we find that the expected 
K-correction is less than 0.4 mag.  Since the exact absolute magnitudes are 
not critical for our study, we will neglect it hereafter.

\section{Spectral Fitting}

The spectra of Type 1 AGNs are generally very complex.  Figure 2 illustrates 
some spectra from our sample.  While in this study 
we are primarily interested in the narrow-line spectrum, which is somewhat 
easier to measure, we will describe in greater detail our method for 
analyzing the entire spectrum.  We illustrate an example of spectral fitting 
in Figure 3.  After removing the Galactic extinction, the spectrum is 
transformed to the restframe using the redshift measured by the SDSS pipeline. 
A full decomposition of the continuum spectrum requires fitting of four 
components.  First, we allow for the presence of a galaxy component.  As 
mentioned in \S2, we expect galaxy contamination to be important in the
lower-luminosity objects in our sample.  While more sophisticated methods of
galaxy modeling have been attempted (e.g., Ho et al. 1993a, 1997a; Greene \&
Ho 2004), Greene \& Ho (2005b) find that a simple scaling and subtraction of a
velocity-broadened K-giant star are usually sufficient for Type 1 AGNs
with the signal-to-noise ratio (S/N) of SDSS.
As in Greene \& Ho (2005b), we use the equivalent width (EW) of the
Ca~{\sc ii} K~ \lamb3934 absorption line as a guide to the level of galaxy
contamination and how much starlight to subtract.  These authors find that
an EW of 1.5 \AA\ for Ca~{\sc ii}~K corresponds to a galaxy contribution of
$\sim 10$\%.  We omit the K-giant component for EW(Ca~{\sc ii}~K) $<$ 1.5 \AA,
and we scale it appropriate for EWs larger than this value.  Second, the 
ultraviolet and optical continuum contains a prominent featureless, nonstellar 
component, which previous studies (e.g., Francis et al. 1991; Vanden Berk
et al. 2001) have approximated by a double power law, with a spectral break at 
$\sim$5000 \AA.  We find, however, that after accounting for the galaxy 
component the second, long-wavelength power-law component is usually 
unnecessary, and thus we model the featureless continuum using a single power 
law.  Third, following Grandi (1982), we include the contribution from a Balmer 
continuum, assuming that it is optically thick and emitted by a uniform 
temperature of $T_e =15,000$ K; we do not attempt to model the higher-order 
Balmer lines.  Lastly, we model 
the ``pseudo-continuum'' generated by the plethora of broad and blended 
Fe~{\sc ii} multiplets, which, for our present application, are particularly 
troublesome for the region containing the \hb\ and \oiii\ emission lines.  
Following standard practice (Boroson \& Green 1992), we remove the Fe blends 
by scaling and broadening an Fe template generated from the spectrum of 
I~Zw~1, kindly provided by T. A. Boroson.  (The region between 3082 \AA\ and 
3685 \AA\ is not modeled because of a gap in the current Fe template.)
After subtracting the galaxy component, we simultaneously fit the quasar 
spectrum with the power-law continuum, Balmer continuum, 
and Fe template.  We perform the fit over the following regions, which are
devoid of strong emission lines: 3550--3645, 4170--4260, 4430--4770, 
5080--5550, 6050--6200, and 6890--7010 \AA. 

With a pure emission-line spectrum at hand, the next task is to fit the 
lines (see bottom of Fig. 3).  Since the continuum near $\sim 3700$ \AA\ 
is still not completely modeled (we neglected higher-order Balmer lines in 
the Balmer continuum model and the Fe template is incomplete), we have decided 
to fit \oii\ \lamb 3727 with a single Gaussian above a locally defined 
continuum.  This procedure, although crude, should be adequate for our purposes.
We note that Greene \& Ho (2005a) find that \oii, at the 
S/N of SDSS, is usually well-fitted by a single 
Gaussian.  However, because of the strong confusion with the broad-line 
region, the lines around \hal\ and \hb---and especially the narrow components 
of these lines themselves---require more complex treatment.  Previous 
studies (Ho et al. 1997b; Greene \& Ho 2004, 2005b) find that the 
\sii\ \lamb\lamb 6716, 6731 doublet serves as a very effective template to 
model \nii\ \lamb\lamb 6548, 6583 and the narrow component of \hal.  
In objects with sufficiently strong \sii\ (S/N \gax\ 3), we fit each of the 
\sii\ lines with a multi-Gaussian model (usually just two suffice), holding 
fixed the known separation of the two lines.  This \sii\ template is then 
used as a model for \nii\ and narrow \hal.  The separation of the \nii\ 
doublet is held fixed, as is its flux ratio (2.96).  In practice, we often 
find it necessary to also fix the position of the \hal\ component relative to 
\nii.  In cases where \sii\ has S/N \lax\ 3, we have no alternative but to use 
\oiii\ \lamb5007 as a template, even though we know that this is not entirely 
adequate because \oiii\ often has a blue, asymmetric wing that is generally 
absent from \sii\ (Greene \& Ho 2005a).  As in Greene \& Ho (2005a, 2005b), we 
model each of the \oiii\ \lamb\lamb 4959, 5007 lines with a two-component 
Gaussian, and we constrain the narrow component of \hb\ to the narrow 
component of \hal, keeping the flux ratio fixed to the Case B$^{\prime}$ value 
of \hal/\hb\ = 3.1 (Osterbrock 1989).\footnote{Note that this constraint 
effectively assumes that the NLR experiences negligible internal extinction.  
We have experimented with leaving free the amplitude of the narrow component 
of \hb.  The resulting distribution of Balmer decrement is sharply peaked at a 
value of \hal/\hb\ = 3.3, suggesting that any internal extinction is evidently
quite modest ($A_V \approx 0.2$ mag).  However, a minority of the objects have 
extremely tiny, unphysical \hal/\hb\ ratios, resulting from severe 
overestimation of the \hb\ strength.  Since the majority of the objects seem 
to be well characterized by small amounts of internal extinction in their NLR, 
we have decided that it would be prudent to fix the \hal/\hb\ ratio to the 
Case B$^{\prime}$ value.  As an aside, it is interesting, and probably not 
coincidental, that Greene \& Ho (2005b) find that the Balmer decrements of the 
broad-line region in their sample of SDSS Type 1 AGNs also indicate little,
if any, internal extinction along the line of sight.} The broad components of 
\hal\ and \hb\ are fitted with multiple Gaussian components.

The vast majority ($\sim$95\%) of our sample yielded unambiguous detection of 
the strongest emission lines considered in this study (\oii, \oiii, \hal, and 
\nii).  The success rate, however, was lower for \hb\ (66\%), and it was 
disappointingly low (44\%) for \oi\ and \sii, which, in addition to being 
weaker, also lie in the reddest portion of the SDSS bandpass, where the 
effective S/N is decreased by systematic noise due to imperfect removal of 
night-sky emission, telluric absorption, and detector fringing.  We set a 
3 $\sigma$ upper limit for the nondetections by assuming that the line has a 
Gaussian profile equal to that of the average of the detected narrow lines and 
an amplitude 3 times the local rms of the local continuum.

\begin{figure*}
\psfig{file=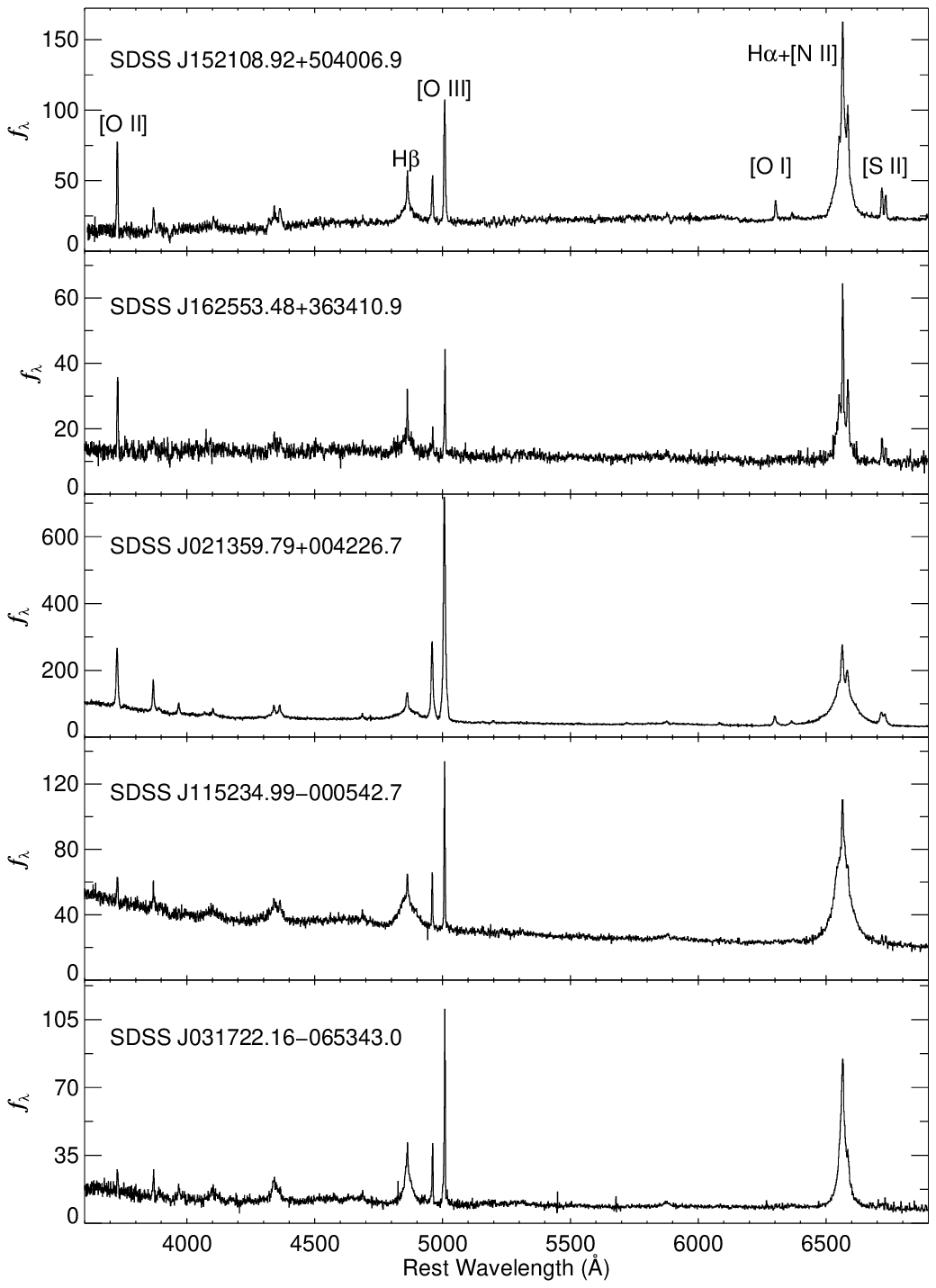,width=18.5cm,angle=0} \figcaption[fig2.ps] {
Sample spectra of our objects, chosen to span a range in S/N and \oii\
strength.  The restframe flux, $f_{\lambda}=f_{\lambda, {\rm obs}}(1+z)^3$,
is in units of $10^{-17}$ ergs s$^{-1}$ cm$^{-2}$ \AA$^{-1}$.
\label{fig2}}
\end{figure*}

\begin{figure*}
\psfig{file=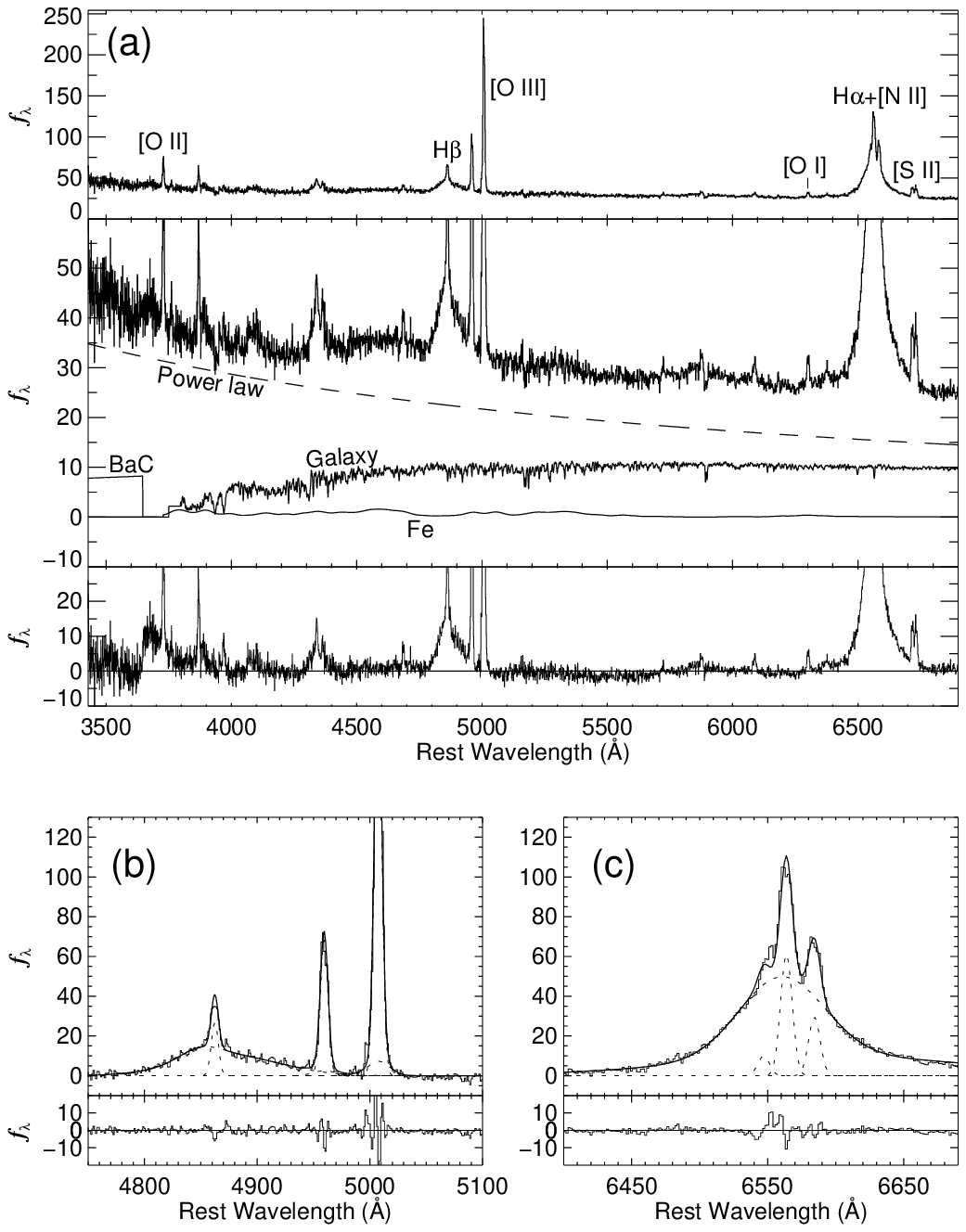,width=18.5cm,angle=0} \figcaption[fig3.ps] {
Example of spectral decomposition for one of the sample objects
(SDSS J161141.95+495847.9). The restframe flux,
$f_{\lambda}=f_{\lambda, {\rm obs}}(1+z)^3$, is in units of $10^{-17}$ ergs
s$^{-1}$ cm$^{-2}$ \AA$^{-1}$. (\textit{a}) The top panel shows the original
spectrum in full scale.  The middle panel gives a magnified view of the
original spectrum, with some
key emission lines labeled, along with the different continuum components
included in the fit [featureless power law, Balmer continuum (BaC), galaxy
(K-giant star), and the Fe template].  The continuum-subtracted emission-line
spectrum is shown in the bottom panel.  Note that the Fe template does not
cover the region between 3082 and 3685 \AA, and the BaC component does not
properly model the higher-order Balmer lines in the region 3650--3800 \AA,
resulting in strong residuals in the continuum near \oii.  To illustrate the
detailed fits to the emission lines, we show an expanded view of the
regions surrounding (\textit{b}) \hb\ and \oiii\ \lamb\lamb4959, 5007 and
(\textit{c}) \hal\ and \nii\ \lamb\lamb6548, 6583.
\label{fig3}}
\end{figure*}

\begin{figure*}
\vskip 5mm
\psfig{file=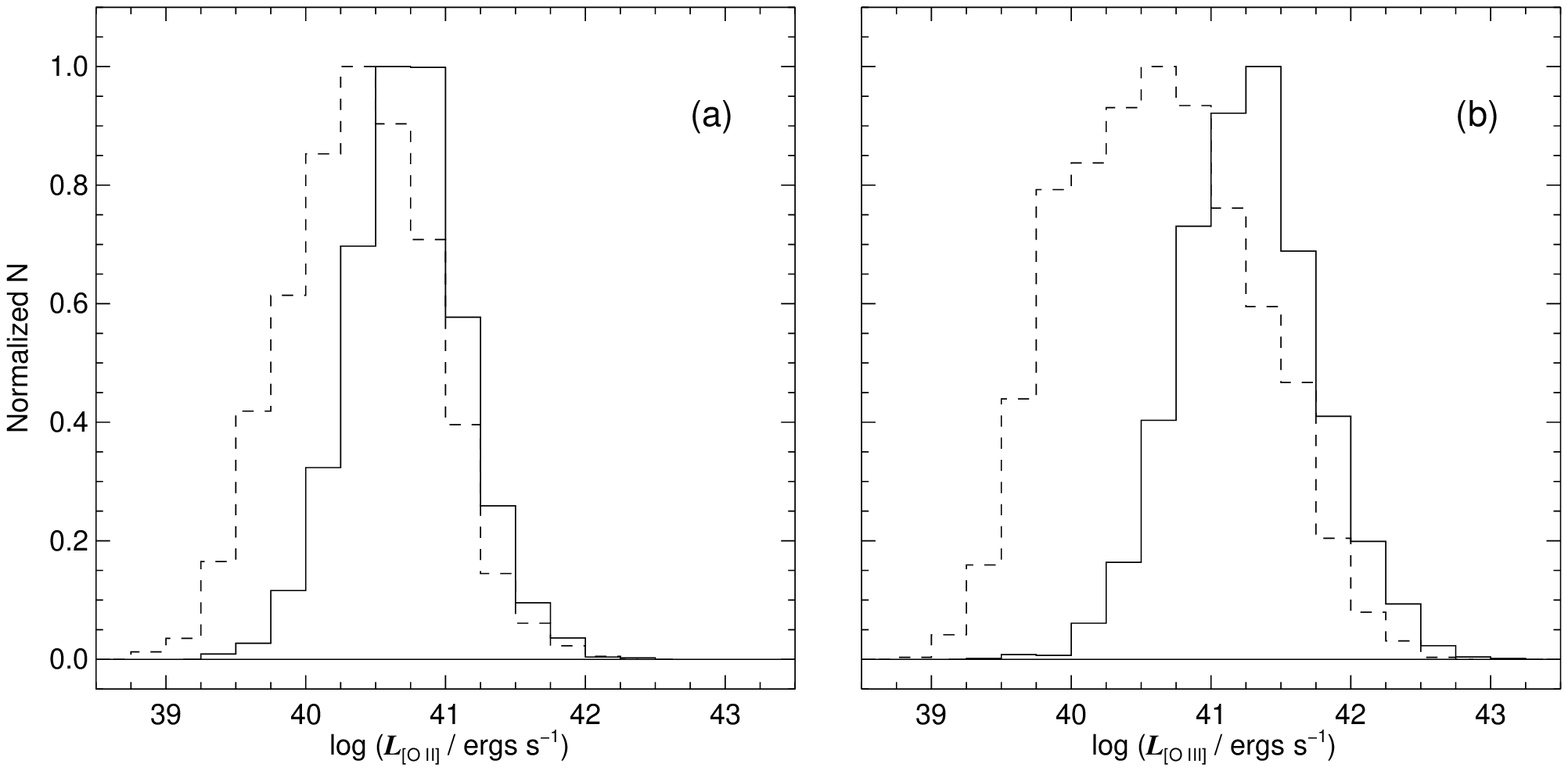,width=18.5cm,angle=0} \figcaption[fig4.ps] {
Relative distribution of (\textit{a}) \oii\ and (\textit{b}) \oiii\
luminosity.  The width of each bin is 0.25 dex.  The solid line represents the
Type 1 AGNs from our sample, and the dashed line represents the Type 2 AGNs
from Kauffmann et al. (2003).  The Type 2 objects have been corrected for
internal extinction, as given in Kauffmann et al.
\label{fig4}}
\end{figure*}
\section{Results}

\subsection{\oii\ and \oiii\ Luminosities}

Figure 4 shows the distributions of \oii\ $\lambda$3727 and 
\oiii\ $\lambda$5007 luminosities for our sample.  The \oii\ luminosities 
range from $\sim 10^{40}$ to $3\times 10^{42}$ ergs s$^{-1}$, with an average 
of $\langle L_{\rm [O~{\sc II}]} \rangle = 5.2 \times 10^{40}$ \lum.  For 
comparison, we overplot the sample of Type 2 AGNs studied by Kauffmann et al. 
(2003), which, in terms of narrow-line emission, is approximately a factor of 
2 less luminous than our sample.  As discussed in Ho (2005), 
this level of \oii\ emission, taking conservative assumptions about AGN 
contamination and corrections for extinction and metallicity, corresponds 
to SFRs of only $\sim$0.5  \solmass\ \peryr.  Figure 5 shows the luminosity of
\oiii\ plotted versus the line ratio \oii/\oiii.  The \oii/\oiii\ ratio spans a
wide range, in the extreme by nearly 2 orders of magnitude from $\sim 0.06$ 
to $\sim 5$, but most cluster around $\sim 0.2-1$.  Assuming that \oiii\ is 
produced entirely by the AGN and is a good tracer of AGN power (e.g., 
Kauffmann et al. 2003), it is clear that the \oii/\oiii\ ratio decreases 
strongly with increasing \oiii\ strength.  For comparison, we also overplot 
the sample of Kauffmann et al. (2003) to illustrate that the lower-luminosity 
Type 2 objects also obey roughly the same pattern.  This trend is easy to 
explain in the context of AGN photoionization models, in which the \oii/\oiii\ 
ratio varies strongly as a function of
\epsfig{file=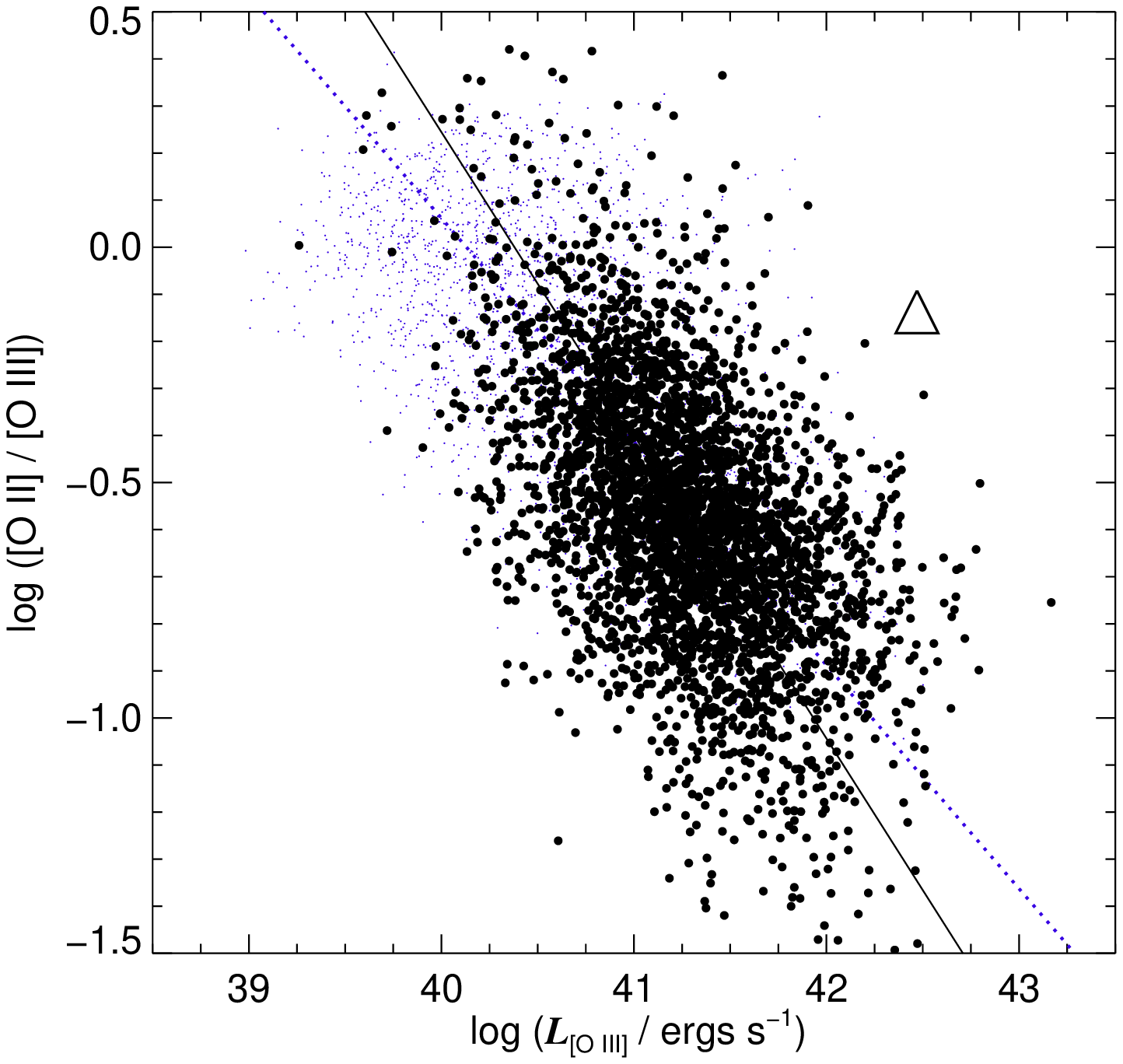,width=8.5cm,angle=0}
\figcaption[fig5.ps] {
The \oii$/$\oiii\ ratio vs. \oiii\ luminosity for Type 1 AGNs (large black
dots) and Type 2 AGNs (small blue dots; from Kauffmann et al. 2003). The solid
and dotted lines give the ordinary least-squares bisector regression for the
Type 1 and Type 2 objects, respectively.  The large triangle marks the
location of the sample of Type 2 quasars from Zakamska et al. (2003), using
the average value of the \oiii\ luminosity and the \oii/\oiii\ ratio measured
from their composite spectrum.  The values from Kauffmann et al. and
Zakamska et al. have been corrected for internal extinction.
\label{fig5}}
\vskip 5mm
of the ionization parameter\footnote{The
ionization parameter, denoted by $U$, is defined to be the ratio of the 
ionizing photon density to the density of hydrogen.} (see below), but it is
difficult to reconcile in a picture wherein the strength of star formation
increases with increasing AGN activity.  In such a scenario, the 
\oii/\oiii\ ratio would either remain constant or rise with increasing 
$L_{\rm [O~{\sc III}]}$.

\subsection{Photoionization Model}

To better constrain the origin of the narrow-line spectrum in general, and of 
the \oii\ line in particular, we performed a new set of photoionization 
calculations using the code CLOUDY (version 94.00; Ferland et al. 1998). 
For an optically thick cloud of a given density and elemental abundance 
illuminated by an input ionizing spectrum, CLOUDY solves the equations of 
statistical and thermal equilibrium and produces a self-consistent model of 
the run of temperature and ionization as a function of depth into the cloud. 
For simplicity, we assume that the cloud has a uniform density, has a 
plane-parallel geometry, and is dust-free. We calculated a grid of models by 
varying three main parameters: the hydrogen density ($10^2$ cm$^{-3}$ 
$\leq n \leq$ $10^6$ cm$^{-3}$), the ionization parameter ($10^{-4} \leq U 
\leq 10^{-1}$), and the shape of the ionizing continuum ($\alpha = -0.5, -1, 
-1.5, -2$, and $-2.5$).  The ionizing continuum has the basic shape described 
by Ho et al. (1993b), where the variation is parameterized by a power-law 
index $\alpha$ between 10 $\mu$m and 50 keV.  We adopt the solar abundances as 
given in Grevesse \& Anders (1989), with the exception of nitrogen whose 
abundance is increased to twice the solar value, as suggested by 
Storchi-Bergmann (1991).

\begin{figure*} \vskip -5mm
\psfig{file=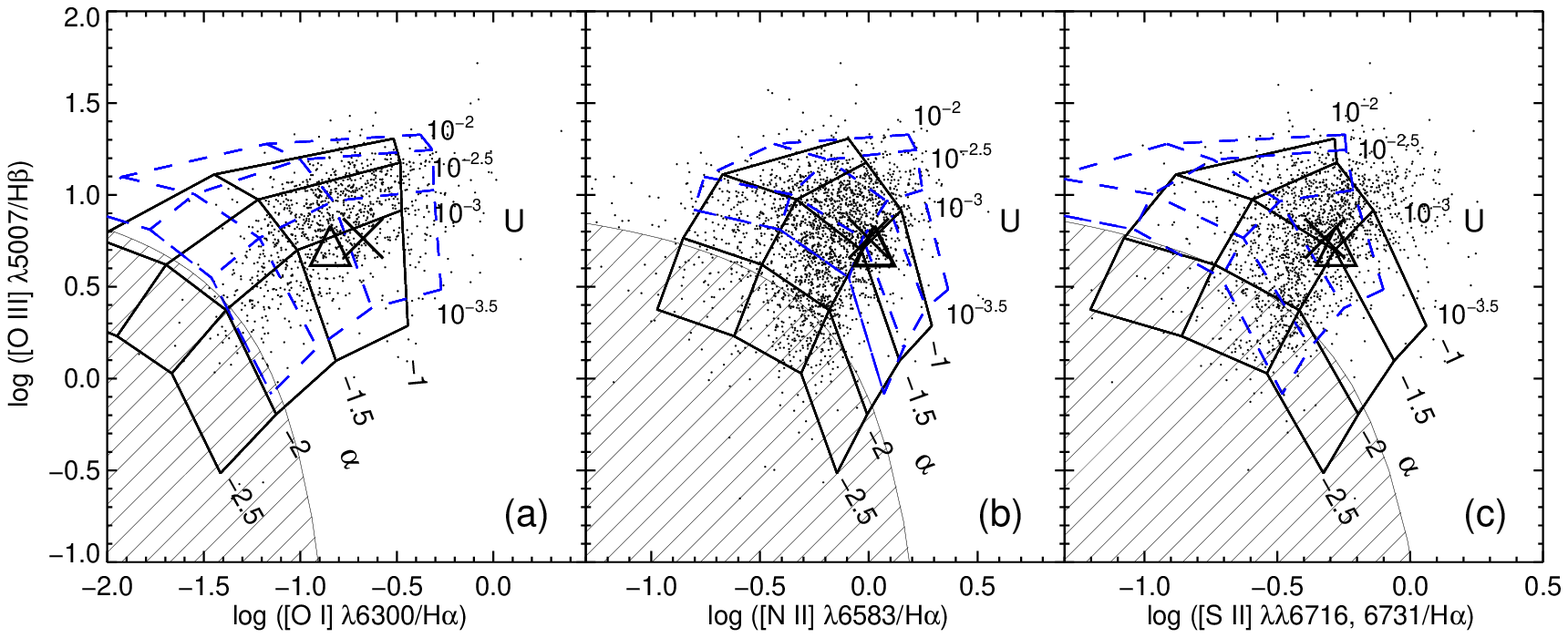,width=18.5cm,angle=0} \figcaption[fig6.ps] {
Line ratios of Type 1 AGNs (small dots) and Type 2 quasars (large triangle,
from composite spectrum of Zakamska et al. 2003) compared with photoionization
models.  The large cross marks the location of the average value of the upper
limits in our sample, which are consistent with the detections.  The three
Veilleux \& Osterbrock (1987) diagrams plot  \oiii\ \lamb5007$/$\hb\ vs.
(\textit{a}) \oi\ \lamb6300$/$\hal, (\textit{b}) \nii\ \lamb6583$/$\hal, and
(\textit{c}) \sii\ \lamb\lamb6716, 6731$/$\hal.   Two grids are shown,
representing $n=10^2\,{\rm cm}^{-3}$ ({\it solid black line}) and
$n=10^4\,{\rm cm}^{-3}$ ({\it dashed blue line}).  Each grid shows models for
$U = 10^{-2}$, $10^{-2.5}$, $10^{-3}$, and $10^{-3.5}$ ({\it top to bottom})
and $\alpha = -2.5$, $-2$, $-1.5$, and $-1$ ({\it left to right}).
The shaded region represents the locus of H~II regions, as
calculated from the starburst models of Kewley et al. (2001).
\label{fig6}} \vskip +5mm
\end{figure*}

\begin{figure*} \vskip -3mm
\psfig{file=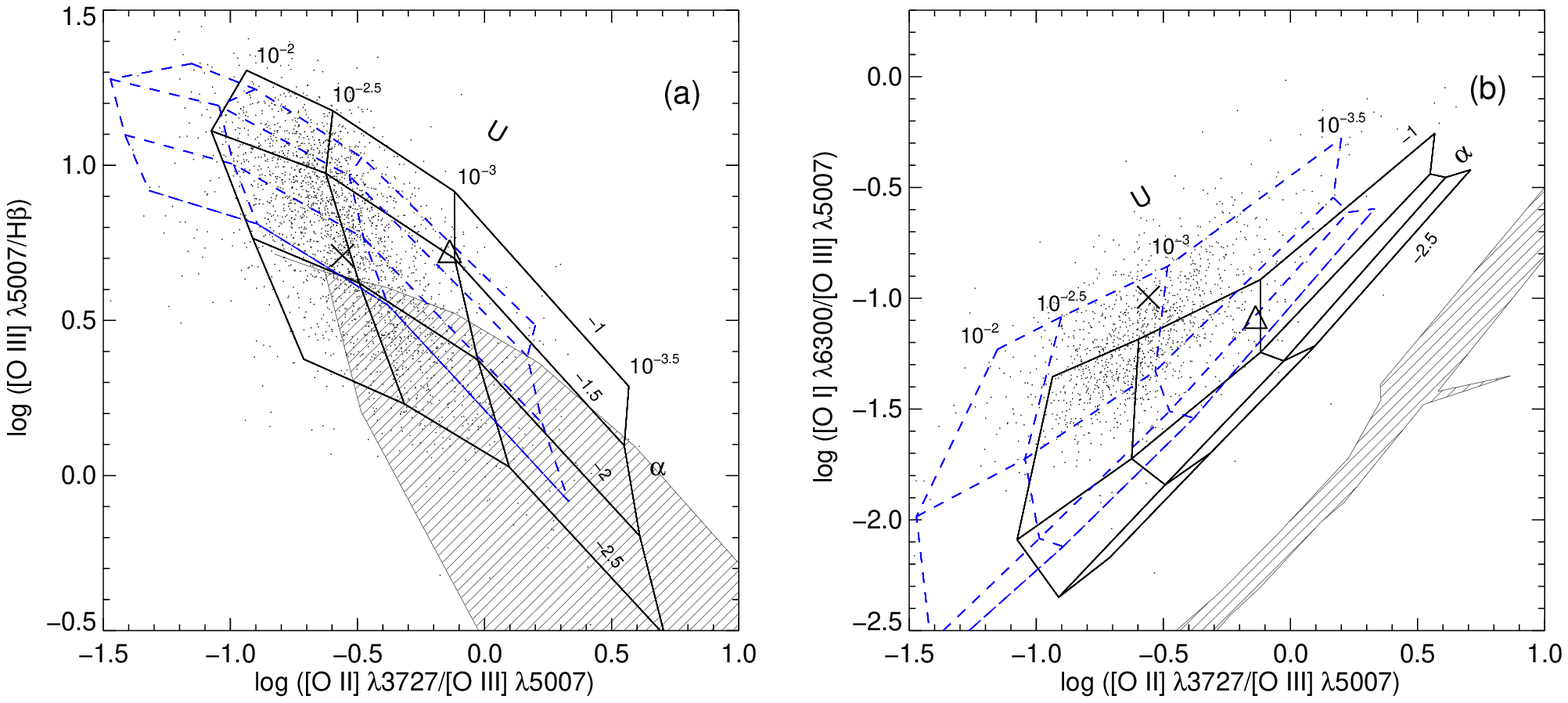, width=18.5cm, angle=0} \figcaption[fig7.ps] {
Line ratios of Type 1 AGNs (small dots) and Type 2 quasars (large triangle,
from composite spectrum of Zakamska et al. 2003, with internal extinction
correction applied) compared with photoionization models.  The large cross
marks the location of the average value of the upper limits in our sample,
which are consistent with the detections.  Diagnostic diagrams showing
(\textit{a}) \oiii\ \lamb5007$/$\hb\ vs.  \oii\ \lamb3727$/$\oiii\
\lamb5007 and (\textit{b}) \oi\ \lamb6300$/$\oiii\ \lamb5007 vs.  \oii\
\lamb3727$/$\oiii\ \lamb5007.  Two grids are shown, representing
$n=10^2\,{\rm cm}^{-3}$ ({\it solid black line}) and $n=10^4\,{\rm cm}^{-3}$
({\it dashed blue line}).  Each grid shows models for $U = 10^{-2}$,
$10^{-2.5}$, $10^{-3}$, and $10^{-3.5}$ ({\it left to right}) and $\alpha =
-2.5$, $-2$, $-1.5$, and $-1$ ({\it bottom to top}).
The shaded region represents the locus of H~II regions, as
calculated from the starburst models of Kewley et al. (2001).
\label{fig7}}
\vskip +5mm
\end{figure*}


We begin by examining our sample using the three diagrams proposed by 
Veilleux \& Osterbrock (1987), which have the advantage of being insensitive 
to extinction corrections.  As can be seen in Figure 6, the majority of 
the sample of Type 1 objects is well covered by our grid of models, spanning 
the following range of parameters: $-2.0 \leq \alpha \leq -1.0$, $10^{-3.3} 
\leq U \leq 10^{-2.0}$, and $10^2\,{\rm cm}^{-3} \leq n \leq 
10^4\,{\rm cm}^{-3}$.  The density estimate is not consistent from one 
diagram to another because the NLR very likely has a wide range of densities, 
and the various forbidden transitions have different critical densities (e.g., 
Filippenko \& Halpern 1984).  Nonetheless, the above range of parameters 
corresponds well to previous estimates made using the same version of CLOUDY 
(Nagao et al.  2001).  

Having established that our sample of Type 1 objects has NLR properties 
that can be readily accommodated using standard AGN photoionization, we next 
turn to two diagnostic diagrams that highlight the \oii\ line.  Figure 7 
compares the line ratio \oii\ \lamb 3727/\oiii\ \lamb5007, which is a sensitive 
indicator of the ionization parameter, to \oiii\ \lamb5007/\hb\ and \oi\ 
\lamb6300/\oiii\ \lamb5007\ (Baldwin et al. 1981; Ferland \& Netzer 1983).  
These diagrams have the advantage of being relatively insensitive to 
abundance, although \oii/\oiii\ is affected by extinction.  The basic 
conclusion from this figure is that the observed line ratios, and in 
particular the \oii\ strength, can be well reproduced with AGN photoionization 
with roughly the same range of parameters previous deduced from the Veilleux 
\& Osterbrock diagrams (Fig. 6).   In other words, there is no need to invoke 
any additional source apart from the AGN, such as star formation, to account 
for the observed range of \oii\ emission.

To reinforce our conclusion that star formation makes a negligible contribution
to the narrow-line spectrum of our objects, we superimpose on Figures 6 and 7 
the theoretical photoionization models of H~II regions calculated by 
Kewley et al. (2001).  These models explore a range of input parameters meant 
to represent realistic conditions encountered in starburst galaxies.  While the
\oii/\oiii\ and \oiii/\hb\ ratios of H~II regions and our sample of AGNs 
overlap substantially (Fig. 7{\it a}), the starburst models, even under the 
most extreme conditions, cannot account for the strength of the low-ionization 
transitions (\nii, \sii, and especially \oi) observed in the AGNs.

\section{Discussion}

\subsection{AGN Feedback}

Ho (2005; see also Ho 2006) recently proposed that the 
\oii\ \lamb3727 emission line can 
effectively trace ongoing star formation in the host galaxies of luminous, 
high-ionization AGNs such as Seyferts and quasars.  The \oii\ line has long 
been used as a SFR indicator in spectroscopic surveys of distant galaxies, but 
previously it had not been applied to systems containing active nuclei.  From a 
review of the extant literature on \oii\ measurements in quasars, Ho came to 
the conclusion that the SFRs in their host galaxies are quite modest, being 
no greater than $\sim 1-10$ \solmass\ \peryr.  This result was somewhat 
unexpected considering the recent evidence for there being a substantial 
post-starburst population in AGNs (Kauffmann et al. 2003).  Perhaps even more
surprising, Ho found, from a more limited sample of low-redshift quasars with 
available CO measurements, that not only are the SFRs low but that the star 
formation efficiencies (SFR per unit gas mass) are also low, suggesting that 
the AGN somehow {\it inhibits}\ star formation.  

Apart from the small sample of low-redshift objects that had individual 
spectroscopic data, Ho's more general conclusions regarding the quasar 
population as a whole was based on published measurements of \oii\ strengths
derived from composite, or statistically averaged, spectra.   To place these
results on a more secure footing, we have undertaken a major task to 
homogeneously measure the \oii\ line in a large sample of Type 1 AGNs.  Apart 
from \oii, we also simultaneously measured a number of other diagnostically 
important narrow emission lines, and we have generated a new set of 
photoionization models to interpret them.  Our analysis indicates that the 
principal narrow emission lines of our sample, including \oii, can be entirely 
explained by AGN photoionization assuming fairly standard parameters.  There 
is no evidence for any excess \oii\ emission arising from a separate origin, 
such as \hii\ regions.  Thus, the observed \oii\ strengths can be used to 
set a limit on the ongoing SFR in these systems.   However, as discussed by 
Ho, a number of uncertainties complicate the SFR estimates based on the \oii\ 
line.  These include the unknown magnitude of dust extinction, the possible 
influence of metallicity corrections, and precisely
how much of the observed \oii\ strength to attribute to star formation. 
For illustrative purposes, we make three simple assumptions (see  Ho 2005 for 
a more detailed discussion): (1) that the amount of dust extinction in AGN host 
galaxies is comparable to that deduced for moderately actively star-forming 
galaxies ($A_V \approx 1$ mag; e.g., Hopkins et al. 2003); (2) that the 
metallicity is twice solar (e.g., Storchi-Bergmann et al. 1998); and (3) that 
one-third of the observed \oii\ strength comes from \hii\ regions.  For a 
typical \oii\ luminosity of $5.2 \times 10^{40}$ \lum\ (Fig. 4{\it a}), the 
empirical calibration of Kewley et al. (2004) yields a SFR of merely 
0.5 \solmass\ \peryr.  This estimate is consistent with, but even lower 
than, than that given by Ho (2005).  To put this estimate in perspective, 
we note that the SFRs of nearby normal spiral galaxies, including the Milky 
Way, range from $\sim$1 to 3 \solmass\ \peryr\ (Solomon \& Sage 1988).  
The situation in AGNs is thus truly unusual.

Unlike the low-redshift quasars considered by Ho (2005), our current sample
of SDSS AGNs has no available information on its cold gas content.  However, 
the unbiased CO survey by Scoville et al. (2003) shows that optically selected 
quasars, not unlike those studied here, are quite gas-rich.   This is not 
unexpected, considering that the nuclei, by selection, are quite active, 
and therefore accreting at a fairly high rate.  Thus, although the evidence 
is certainly less direct, we can reasonably surmise that the objects studied 
here also may also possess abnormally low star formation efficiencies.

How can this finding be reconciled with Kauffmann et al.'s (2003) result that 
luminous AGNs show a preponderance of post-starburst activity?  A possible 
solution is to invoke an evolutionary scenario, not unlike that proposed 
by Sanders et al. (1988), in which a starburst phase {\it precedes}\ the 
optically revealed quasar phase.  From the age estimates given in Kauffmann 
et al., the time lag seems to be approximately $10^8$ to $10^9$ years.  The 
starburst phase terminates not because all of the gas has been transformed 
into stars, nor because it has been completely expelled by AGN or supernova 
feedback, since evidently plenty of gas coexists during the quasar phase, but 
because the active nucleus somehow prevents the gas from forming stars. 
Through a combination of radiative heating and kinetic energy injection, 
an AGN is expected to significantly alter the thermodynamical state of 
the gas in the circumnuclear regions of galaxies (Begelman 1993; Di~Matteo et 
al. 2005), but precisely how this results in suppression of star formation
remains to be elucidated.

\begin{figure*}[t]
\centerline{\psfig{file=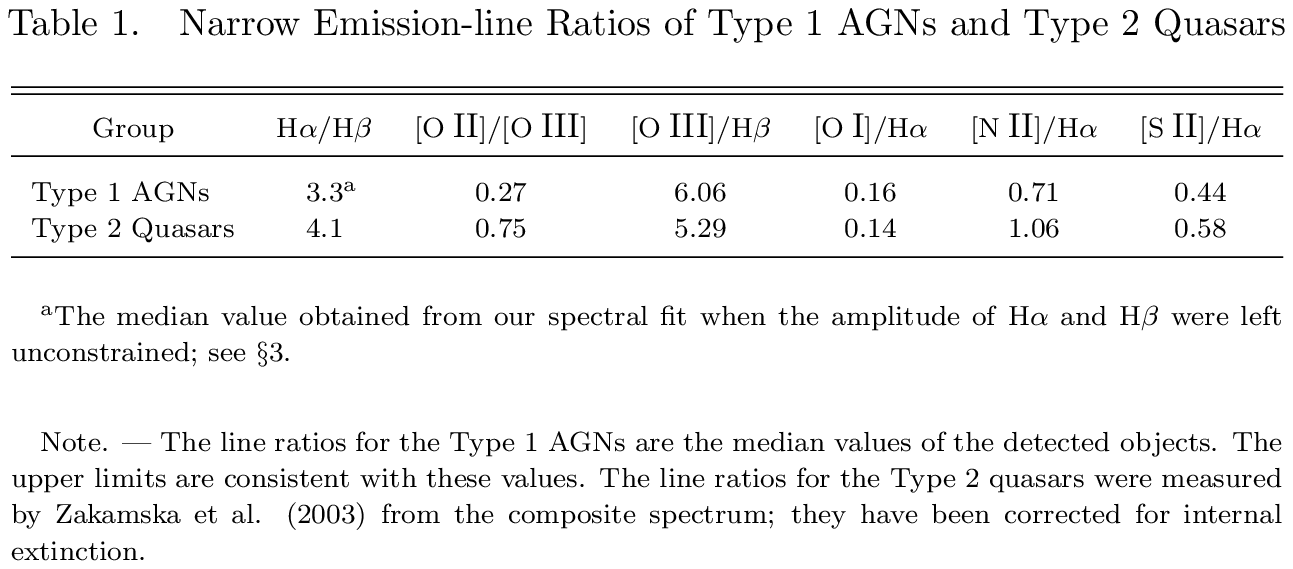,keepaspectratio=true,angle=0}}
\end{figure*} 
\vskip 5mm

\subsection{Possible Evidence for Starburst Activity in Type 2 Quasars}

Zakamska et al. (2003) have drawn attention to a class of narrow-line AGNs 
from the SDSS whose \oiii\ line strength, when translated into continuum 
luminosities, place them in the league of quasars.  This, along with the 
detection in some sources of polarized broad-line emission through 
spectropolarimetry (Zakamska et al. 2005) and heavy photoelectric absorption 
through X-ray observations (Ptak et al. 2006), has prompted the identification 
of these objects with the long-sought ``Type 2 quasars'' anticipated from the
the basic orientation-dependent unification model of AGNs.   If the 
Zakamska et al. objects truly are intrinsically identical to normal, 
high-luminosity Type 1 AGNs, but simply viewed from a line of sight 
obscured from the nucleus, then the narrow-line spectrum of the two groups, 
assumed to be isotropic, should be similar.  We were surprised to discover 
that this appears not to be the case.  Comparison of the line strengths
for our Type 1 objects with those tabulated for the composite SDSS spectrum of 
Zakamska et al. (2003) reveals that both groups have similar line intensity 
ratios, with two noticeable exceptions (see Table 1).  First, the Type 2 
sources have a higher Balmer decrement, and hence a higher inferred internal 
extinction for their NLR, compared to the Type 1 sources. 
Recall from our spectral fitting (\S3) that, when allowed to vary freely, the 
majority of our sources yielded an \hal/\hb\ ratio of 3.3, to be compared with 
\hal/\hb\ = 4.1 for the Type 2 composite.  If interpreted in terms of 
dust extinction assuming an intrinsic Balmer decrement of 3.1 and a 
Galactic extinction curve, this translates into an internal extinction of 
$A_V = 0.2$ mag versus $A_V = 0.9$ mag, respectively.
Second, and more pertinent to the central theme of this study, we find that 
the Type 2 quasars exhibit an anomalously enhanced (extinction 
corrected) \oii/\oiii\ ratio.  This point is illustrated in Figures 6 and 7, 
where we highlight the location of the Zakamska composite in relation to the 
locus of the Type 1 sources.  While the relatively high \oii/\oiii\ ratio can 
normally be attributed to a somewhat lower ionization parameter, we note that 
this explanation is in serious conflict with the high luminosity of these
sources.  As discussed in \S 4.2, the \oii/\oiii\ ratio varies strongly with 
\oiii\ luminosity.  For $L_{\rm [O~{\sc III}]} \approx 3\times 10^{42}$ \lum, 
the average for the Zakamska et al. sample, Figure 5 predicts that the 
\oii/\oiii\ ratio should be $\sim 0.06-0.1$.  By contrast, the Type 2 quasar
composite gives \oii/\oiii\ = 0.75, nearly an order of magnitude larger than 
the predicted value.  We speculate that the enhanced \oii\ emission in the 
Type 2 quasars comes from star formation.  If the observed \oii/\oiii\ ratio 
in Type 2 quasars is $\sim$10 times larger than expected for their \oiii\ 
luminosity, then the excess \oii\ luminosity of $\sim 2\times 10^{42}$ \lum, 
adopting the assumptions specified in \S5.1, corresponds to an estimated SFR 
of $\sim 20$ \solmass\ \peryr.  This level of star formation qualifies as a 
moderately healthy starburst.  Incidentally, we note that a dusty, gas-rich 
environment characteristic of starbursts may naturally account for the higher 
internal extinction mentioned above.

The possible existence of significant ongoing star formation in Type 2 
quasars, in contrast to their stark absence in normal (Type 1) quasars, 
presents an obvious challenge to the conventional orientation-dependent model 
of AGN unification (Antonucci 1993).  Type 2 objects are not intrinsically the 
same as Type 1 objects, but instead may represent an earlier evolutionary 
phase, perhaps similar to the scenario discussed in \S5.1, wherein the 
central engine is either not yet fully developed or else lies buried by 
extended, dusty regions of star formation.  In this context, the optically 
selected Type 2 quasars closely resemble those ultraluminous infrared galaxies 
that host both a strong starburst and a highly obscured AGN (e.g., Genzel 
et al. 1998; Gerssen et al. 2004).

Observations of radio-loud AGNs have suggested that the \oiii-emitting 
region, unlike that producing \oii, may not be entirely isotropic (Jackson \& 
Browne 1990; Hes et al. 1996). If this is generically true of all AGNs, 
then the \oii/\oiii\ ratio is orientation-dependent, and it may account for 
the higher ratios in Type 2 quasars compared to Type 1 quasars.  This 
explanation, however, is unsatisfactory because Figure 5 shows no evidence that 
this pattern holds for lower-luminosity sources.

\section{Summary}
We present a detailed analysis of the narrow-line spectrum of a large, 
homogeneous sample of broad-line (Type 1) AGNs selected from SDSS, with the 
primary aim of using the strength of the \oii\ \lamb3727 line to constrain 
the ongoing star formation rate in the host galaxies.  Comparing the 
measurements with a new set of photoionization models calculated using the 
code CLOUDY, we find that the principal narrow optical lines, including 
\oii, can be readily reproduced using conventional AGN photoionization 
parameters.  This places strong constraints on any additional starburst
contribution to the \oii\ strength, with resulting limits on the inferred star 
formation rate.  Consistent with the recent study of Ho (2005), the host 
galaxies of Type 1 AGNs evidently experience very modest ongoing star 
formation.  On the other hand, the sample ``Type 2'' quasars identified from 
SDSS (Zakamska et al. 2003) exhibits enhanced \oii\ emission, which we argue 
arises from significant ongoing starburst activity.  We propose an evolutionary 
scenario that can account for these observations.

\acknowledgements 
We would like to thank T. A. Boroson for sending us the I Zw 1 iron 
template, G. J. Ferland for making available the code CLOUDY, L. J. Kewley for
making available her model calculations of H~II regions, and an anonymous 
referee for a helpful critique of our work.
We are grateful to the entire SDSS collaboration for providing access to their 
invaluable database.  M.K. has been supported in part by the BK 21 program.
M.K. and M.I. acknowledge the support from the grant No. R01-2005-000-10610-0 
provided by the Basic Research Program of the Korea Science \& Engineering
Foundation.  The work of L.C.H.  was supported by the Carnegie 
Institution of Washington and by NASA through grants from the Space Telescope 
Science Institute (operated by AURA, Inc., under NASA contract NAS5-26555).


\begin{thebibliography}{}
\bibitem[]{}
Abazajian, K., \etal\ 2005, \aj, 129, 1755

\bibitem[]{}
Antonucci, R. 1993, \annrev, 31, 473

\bibitem[]{}
Baldwin, J. A., Phillips, M. M., \& Terlevich, R. 1981, PASP, 93, 5

\bibitem[]{}
Begelman, M. C. 1993, in The Environment and Evolution of Galaxies,
ed. J. M. Shull \& H. A. Thronson Jr. (Dordrecht: Kluwer), 369

\bibitem[]{}
Begelman, M.~C., \& Nath, B. B. 2005, \mnras, 361, 1387

\bibitem[]{}
Blanton, M. R., et al. 2003, \apj, 592, 819

\bibitem[]{}
Boisson, C., Joly, M., Moultaka, J., Pelat, D., \& Serote Roos, M. 2000,
\aa, 357, 850

\bibitem[]{}
Boroson, T. A., \& Green, R. F. 1992, \apjs, 80, 109

\bibitem[]{}
Canalizo, G., \& Stockton, A. 2000, \apj, 528, 201

\bibitem[]{}
Cardelli, J. A., Clayton, G. C., \& Mathis, J. S. 1989, ApJ, 345, 245

\bibitem[]{}
Cid Fernandes, R., Jr., Heckman, T.~M., Schmitt, H.~R., Golz\'alez Delgado,
R.~M., \& Storchi-Bergmann, T. 2001, \apj, 558, 81

\bibitem[]{}
Coleman, G. D., Wu, C.-C., \& Weedman, D. W. 1980, \apjs, 43, 393

\bibitem[]{}
Di Matteo, T., Springel, V., \& Hernquist, L. 2005, Nature, 433, 604

\bibitem[]{}
Ferland, G.~J., Korista, K.~T., Verner, D.~A., Ferguson, J.~W., Kingdon,
J.~B., \& Verner, E.~M. 1998, \pasp, 110, 761

\bibitem[]{}
Ferland, G. J., \& Netzer, H. 1983, \apj, 264, 105

\bibitem[]{}
Ferland, G. J., \& Osterbrock, D. E. 1986, \apj, 300, 658

\bibitem[]{}
Ferrarese, L., \& Merritt, D. 2000, \apj, 539, L9

\bibitem[]{}
Filippenko A. V., \& Halpern, J. P. 1984, \apj, 285, 458

\bibitem[]{}
Francis, P.~J., Hewett, P.~C., Foltz, C.~B., Chaffee, F.~H., Weymann, R.~J.,
\& Morris, S.~L. 1991, \apj, 373, 465

\bibitem[]{}
Gebhardt, K., et al.  2000, \apj, 539, L13

\bibitem[]{}
Genzel, R., et al.  1998, \apj, 498, 579

\bibitem[]{}
Gerssen, J., van der Marel, R.~P., Axon, D., Mihos, J. C., Hernquist, L.,
\& Barnes, J. E. 2004, \aj, 127, 75

\bibitem[]{}
Grandi, S. A. 1982, \apj, 255, 25

\bibitem[]{}
Greene, J. E., \& Ho, L. C. 2004, \apj, 610, 722

\bibitem[]{}
------. 2005a, \apj, 627, 721

\bibitem[]{}
------. 2005b, \apj, 630, 122

\bibitem[]{}
Grevesse, N., \& Anders, E. 1989, in AIP Conf. Proc. 183,
Cosmic Abundances of Matter, ed. C. J. Washington (New York: AIP), 1

\bibitem[]{}
Haas, M., et al. 2003, \aa, 402, 87

\bibitem[]{}
Haehnelt, M. G., \& Rees, M. J. 1993, \mnras, 263, 168

\bibitem[]{}
Hamann, F., Dietrich, M., Sabra, B. M., \& Warner, C. 2004, in Carnegie
Observatories Astrophysics Series, Vol. 4: Origin and Evolution of the
Elements, ed. A. McWilliam \& M. Rauch (Cambridge: Cambridge Univ. Press), 443

\bibitem[]{}
Hes, R., Barthel, P. D., \& Fosbury, R.~A.~E. 1996, \aa, 313, 423

\bibitem[]{}
Ho, L.~C. 2004, ed., Carnegie Observatories Astrophysics Series, Vol. 1: 
Coevolution of Black Holes and Galaxies (Cambridge: Cambridge Univ. Press)

\bibitem[]{}
------. 2005, ApJ, 629, 680

\bibitem[]{}
------. 2006 (astro-ph/0511157)

\bibitem[]{}
Ho, L. C., Filippenko, A. V., \& Sargent, W. L. W. 1993a, ApJ, 417, 63

\bibitem[]{}
------. 1997a, ApJS, 112, 315

\bibitem[]{}
------. 2003, \apj, 583, 159

\bibitem[]{}
Ho, L.~C., Filippenko, A.~V., Sargent, W.~L.~W., \& Peng, C.~Y. 1997b, \apjs,
112, 391

\bibitem[]{}
Ho, L. C., Shields, J. C., \& Filippenko, A.~V. 1993b, ApJ, 410, 567

\bibitem[]{}
Hopkins, A.~M., et al. 2003, \apj, 599, 971

\bibitem[]{}
Jackson, N., \& Browne, I. W. A. 1990, Nature, 343, 43

\bibitem[]{}
Jahnke, K., et al. 2004, \apj, 614, 568

\bibitem[]{}
Kauffmann, G., et al. 2003, \mnras, 346, 1055

\bibitem[]{}
Kauffmann, G., \& Haehnelt, M. G. 2000, \mnras, 311, 576

\bibitem[]{}
Kennicutt, R.~C. 1998, \annrev, 36, 189

\bibitem[]{}
Kewley, L.~J., Dopita, M.~A., Sutherland. R.~S., Heisler, C. A., \& Trevena, J. 2001, \apj, 556, 121

\bibitem[]{}
Kewley, L.~J., Geller, M.~J., \& Jansen, R.~A. 2004, \aj, 127, 2002

\bibitem[]{}
Magorrian, J., et al.  1998, \aj, 115, 2285

\bibitem[]{}
Martini, P. 2004, in Carnegie Observatories Astrophysics Series, Vol. 1: 
Coevolution of Black Holes and Galaxies, ed. L. C. Ho (Cambridge: Cambridge 
Univ. Press), 170

\bibitem[]{}
Nagao, T., Murayama, T., \& Taniguchi, Y. 2001, \apj, 546, 744

\bibitem[]{}
Osterbrock, D.~E.  1989, Astrophysics of Gaseous Nebulae
and Active Galactic Nuclei (Mill Valley, CA: Univ. Science Books)

\bibitem[]{}
Ptak, A., Zakamska, N. L., Strauss, M. A., Krolik, J. H., Heckman, T. M.,
Schneider, D. P., \& Brinkmann, J. 2006, \apj, in press (astro-ph/0510204)

\bibitem[]{}
Robertson, B., Hernquist, L., Cox, T. J., Di Matteo, T., Hopkins, P. F.,
Martini, P., \& Springel, V. 2006, \apj, submitted (astro-ph/0506038)

\bibitem[]{}
Sanders, D. B., Soifer, B. T., Elias, J. H., Madore, B. F., Matthews, K., Neugebauer, G., \& Scoville, N. Z. 1988, \apj, 325, 74

\bibitem[]{}
Schlegel, D.~J., Finkbeiner, D.~P., \& Davis, M. 1998, \apj, 500, 525

\bibitem[]{}
Scoville, N.~Z., Frayer, D.~T., Schinnerer, E., \& Christopher, M. 2003,
\apj, 585, L105

\bibitem[]{}
Silk, J., \& Rees, M. J. 1998, \aa, 331, L1

\bibitem[]{}
Solomon, P.~M., \& Sage, L.~J. 1988, \apj, 334, 613

\bibitem[]{}
Spergel, D. N., et al. 2003, \apjs, 148, 175

\bibitem[]{}
Storchi-Bergmann, T. 1991, \mnras, 249, 404


\bibitem[]{}
Storchi-Bergmann, T., Schmitt, H.~R., Calzetti, D., \& Kinney, A.~L. 1998,
\aj, 115, 909

\bibitem[]{}
Stoughton, C., \etal\ 2002, \aj, 123, 485

\bibitem[]{}
Strauss, M. A., et al. 2002, \aj, 124, 1810

\bibitem[]{}
Vanden Berk, D. E., et al. 2001, \aj, 122, 549

\bibitem[]{}
Veilleux, S., \& Osterbrock, D. E. 1987, ApJS, 63, 295

\bibitem[]{}
York, D.~G., et al. 2000, \aj, 120, 1579

\bibitem[]{}
Zakamska, N. L., et al. 2003, \aj, 126, 2125

\bibitem[]{}
------. 2005, \aj, 129, 1212

\end{thebibliography}
\end{document}